\documentclass{webofc}
\usepackage[varg]{txfonts}   
\begin{document}
\title{Holography of Strongly Coupled Gauge Theories}
%
%

\author{\firstname{} \lastname{Nick Evans}\inst{1}\fnsep\thanks{\email{evans@southampton.ac.uk}} }
\institute{Department of Physics and Astronomy, University of Southampton, Southampton, SO17 1BJ, UK }

\abstract{%
  The "periodic table" of strongly coupled gauge theories remains only sketchily understood. Holography has developed to the point where bottom up constructions can describe the spectrum of individual gauge theories (based on assumptions of their running) including quarks in different representations and higher dimension operators. I highlight the method with a "perfected" version of an AdS dual of QCD and results for composite higgs models with two representations of quarks. The method raises questions about the degree to which energy scales can be split in generic gauge theories including whether confinement and chiral symmetry breaking are linked.
}
\maketitle
\section{Introduction}
\label{intro}
Asymptotically free gauge theories are a key element of the Standard Model  of particle physics - SU(3) gauge theory describes the strong force and the SU(2) theory the weak force. They likely also play a role beyond the Standard Model and possible examples include technicolour \cite{Hill:2002ap}, composite Higgs models \cite{Cacciapaglia:2020kgq}, strongly coupled dark matter \cite{Kribs:2016cew} or dynamical supersymmetry breaking \cite{Affleck:1983mk}. Despite this our knowledge of the spectrum of the set of asymptotically free theories is rather poor. Lattice studies only exist for a smattering of such theories and there remains debate over issues such as where the edge of the conformal window lies for SU(3) theories with fundamental quarks \cite{Hasenfratz:2019dpr,Fodor:2016zil}. Even for theories that are known to break chiral symmetries the spectrum has frequently not yet been computed. It seems like we should do a better job here of understanding the "periodic table"  of these theories. 

Here I want to show that holographic modelling of these gauge theories \cite{Erdmenger:2020lvq,Erdmenger:2020flu} is a useful tool that can quickly provide an estimate of the spectrum. Although these models are not first principle, they can provide a helpful exchange of ideas with lattice studies that can highlight important questions to address.

\section{Holographic Magic}
\label{holo}

Holography \cite{Maldacena:1997re,Witten:1998qj} is a weakly coupled dual description of strongly coupled gauge dynamics that has emerged from string theory. Formally the dual should be a string theory but for ground states of glueballs and meson  a point particle,  gravitational theory works well  for the dual description  \cite{Erlich:2005qh,DaRold:2005mxj} (it is formally justified in the large $N_c$ limit where the QCD string tension is very high). Full descriptions only exist for a small number of gauge theories close to supersymmetric cases but some generic lessons have emerged that can be used for model building. Let's briefly review three of these pieces of "magic".

\begin{itemize}
\item {\bf Renormalization Group Flow:} RG scale emerges as an extra holographic direction in these descriptions. In a five dimensional Anti-de-Sitter (AdS) space 
\begin{equation} ds^2 = {dr^2 \over r^2} + r^2 dx_{3+1}^2 \end{equation}
The radial direction $r$ is the RG scale and slices at some fixed $r$ describe the gauge theory on the 3+1d sub-space. Fields in the AdS bulk have solutions that describe gauge invariant obervables ie operators ${\cal O}$ and sources ${\cal J}$.

\item {\bf Chiral Symmetry Breaking:} In QCD the chiral flavour symmetry is broken SU(2)$_L \times$SU(2)$_R \rightarrow$SU(2)$_V$ by the formation of a vev for the operator $\bar{u}_L u_R + \bar{d}_L d_R + h.c.$
Holographically \cite{Erdmenger:2007cm} we describe the operator as a dimension one scalar field, $L$, in AdS$_5$ satisfying the Klein-Gordon equation \cite{Alvares:2012kr}
\begin{equation} \partial_r [ r^3 \partial_r L] - r \Delta m^2 L = 0, \hspace{1cm}  L = {m \over r^\gamma} + {\langle \bar{q} q \rangle \over r^{2-\gamma} }, \hspace{0.5cm} \gamma(\gamma -2) = \Delta m^2 \end{equation}
When $\Delta m^2=0$ the solutions describe a dimension one mass, $m$, and the dimension three quark condensate. For non-zero $\Delta m^2$ these operators develop an anomalous dimension $\gamma$ as shown.

In AdS$_5$ there is an instability bound for a scalar called the Breitenlohner Freedman bound \cite{Breitenlohner:1982jf} which here corresponds to $\Delta m^2=-1$. If $\Delta m^2$ passes through this bound then the massless $L=0$ solution is unstable. Interestingly here this corresponds to $\gamma=1$ - if the  dimension of $ \bar{q} q $ falls to two then there is an instability to the generation of a quark condensate. This criteria precisely matches that from gap equation analysis \cite{Cohen:1988sq} and which has been used to motivate the first guesses for the edge of the conformal window \cite{Appelquist:1996dq}. 

We can thus make holographic models of chiral symmetry breaking which are essentially a higgs theory, the scalar field is $L$, but whose potential changes with RG scale (radial direction $r$). The potential ($\Delta m^2$)  is determined by the running of the anomalous dimension which in turn is determined by the gauge dynamics. At the scale where $\gamma=1$ the BF bound is violated and chiral symmetry breaking occurs. 

\item {\bf Transitioning to the IR Meson Theory:} finally holography allows us to determine the low energy mesonic theory. For a given background vacuum, for example, a solution for the field $L$, we can allow infinitesimal fluctuations of the form $\delta(r) e^{ik.x}, ~~k^2=-M^2$ which describe fluctuations of the $\bar{q} q$ operator (the $\sigma$ or $f_0$ state). Generically one obtains a Sturm-Louville system which only has normalizable solutions for particular $M^2$. Other states such as the $\rho$ meson can be included via a gauge field dual to $\bar{q} \gamma^\mu q$ and so forth. This method determines the meson spectrum. Substituting these solutions back into the action and integrating over the radial direction returns a 3+1d theory of the mesons and their interaction couplings. This is the standard AdS/QCD methodology
\cite{Erlich:2005qh,DaRold:2005mxj}. 

\end{itemize}

\subsection{Dynamic AdS/QCD}
\label{DAQ}

The simplest model that incorporates these points is the Dynamic AdS/QCD theory \cite{Alho:2013dka}
\begin{equation} S = \int d^4x ~d \rho ~{\rm Tr}~ \rho^3 \left[ { 1 \over \rho^2 + |X|^2} |D X|^2 + {\Delta m^2 \over \rho^2 }|X|^2 + {1 \over 2 \kappa^2} (F_V^2 + F_A^2)  ~+ \bar{\Psi} \left( \slash \hspace{-0.2cm} {D}_{\text{AAdS}} - m \right) \Psi  \right] \end{equation}
where $X=Le^{2 i \pi^a T^a}$ describes the quark condensate/$\sigma$ and pion fields, $F_V$ the $\rho$ meson and $F_A$ the  $a$ meson. $\Psi$ is a Dirac field that can describe the nucleon. The factors of $X$ in the metric (here $r^2=\rho^2+|X|^2$) are deduced from top down (probe D7 brane) models and are the

\begin{figure}[h]
\centering
\includegraphics[width=6cm]{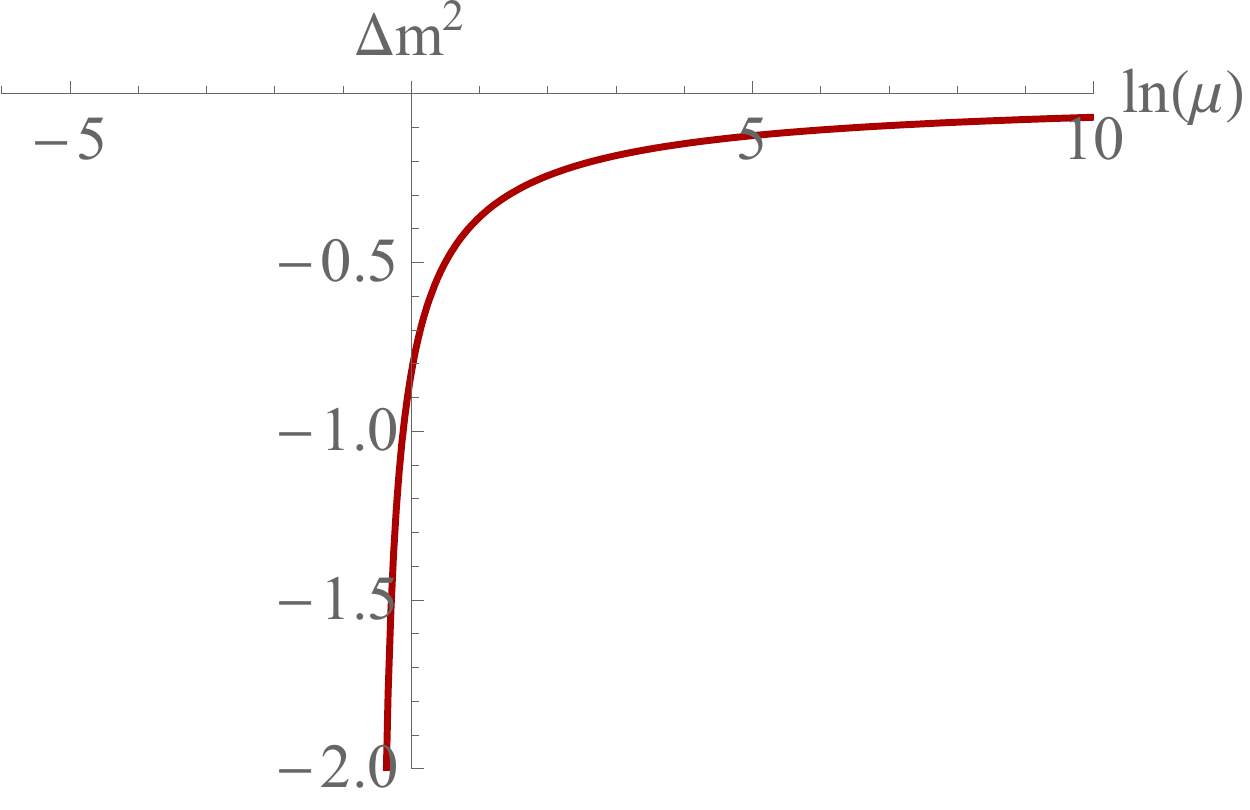}  \hspace{0.5cm}  \includegraphics[width=6cm]{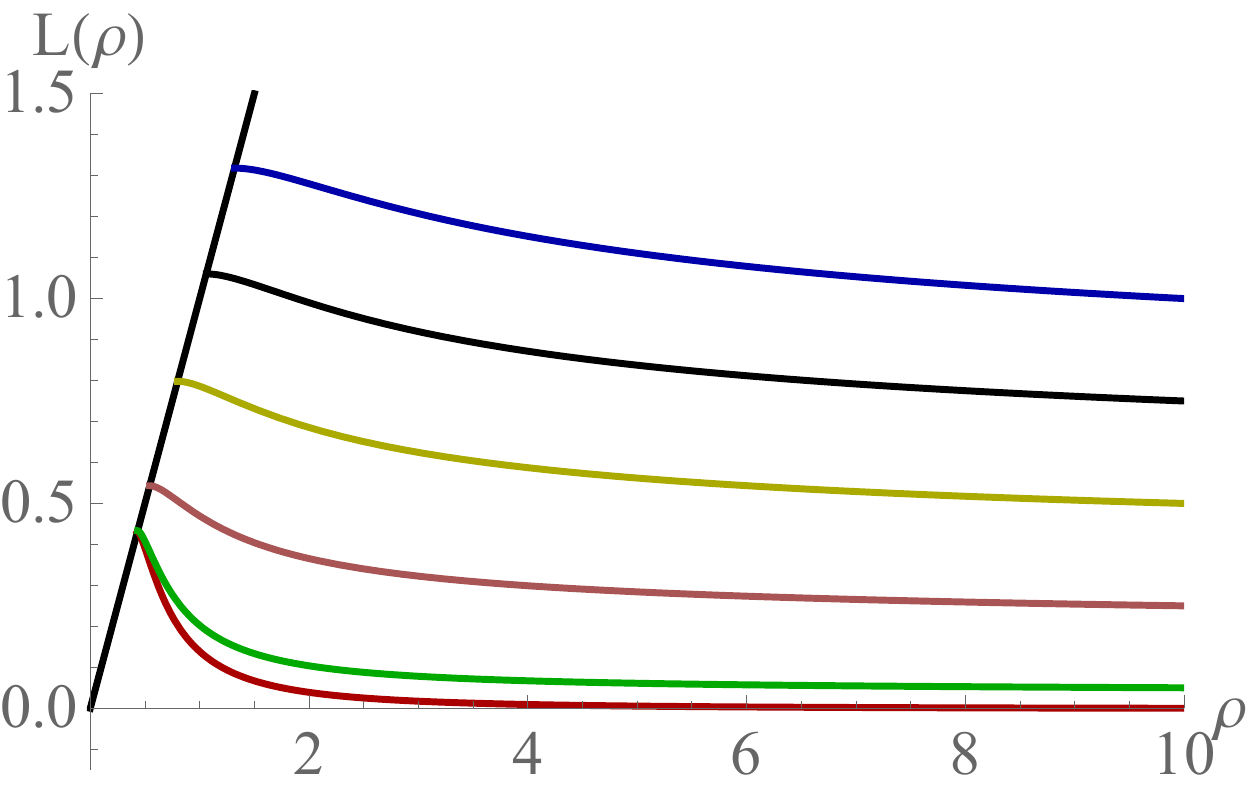}  \\
\includegraphics[width=6cm]{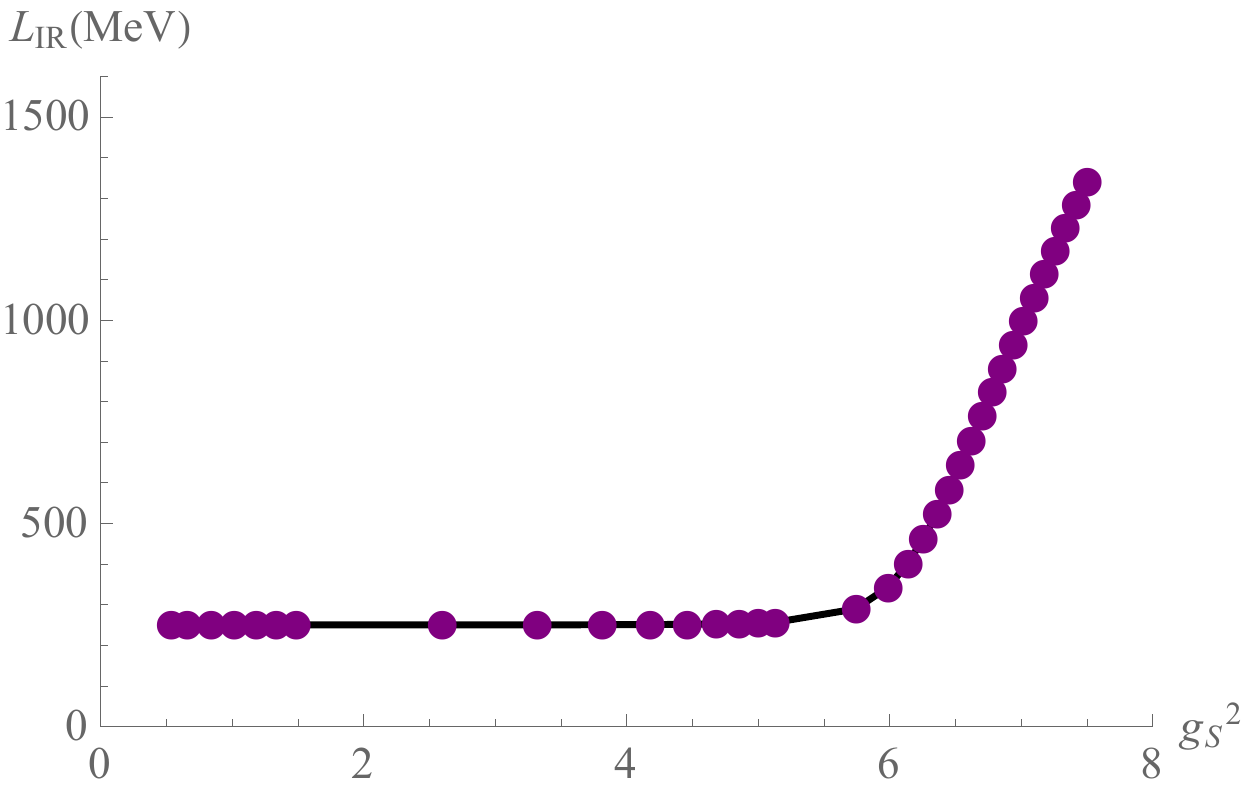}  
\caption{Top left: the running of the mass in the Dynamic AdS/QCD theory for 2 flavour SU(3) QCD. Top right:  the resulting vacuum $L(\rho)$ profiles which asymptote at large $\rho$ to the quark mass. Bottom: the IR quark mass against NJL coupling in two flavour QCD using the solutions above and (\ref{NJL}).}
\label{fig-1}       
\end{figure}

\noindent simplest method to communicate the background quark condensate to the fluctuation fields. 
$\Delta m^2$ can be  input using the perturbative results for the running of $\gamma$ in a theory
\begin{equation} \Delta m^2 = - 2 \gamma = -{3 (N_c^2-1) \over 2 N_c \pi} \alpha \end{equation}
where here $\alpha$ runs according to, for example, the two loop perturbative results with the holographic radial direction $\rho$ directly set to be the RG scale $\mu$ (which we extend into the non-perturbative regime as an ansatz for the running that includes conformal window behaviour \cite{Appelquist:1996dq}). A key benefit of the model is that the only parameters are $N_c, N_f$ and $\Lambda_{QCD}$ as in the true QCD theory. 
More complex models exist – especially note the VQCD theory of Kiritsis and Jarvinen \cite{Jarvinen:2011qe}
 - that attempt to decouple the quarks, include confinement below the chiral symmetry breaking scale, and explain stringy Regge trajectories. These are heroic pieces of work but necessarily contain  more assumptions than this simple model.

The top figures in Figure 1 show the running of $\Delta m^2$ for 2 flavour SU(3) QCD and the resulting profiles for $L$, asymptoting at large $\rho$ to the quark mass $m$.  

\subsection{Higher Dimension Operators (HDOs) }

Holography is a strong weak duality and so the gravitational dual should only describe the IR region of an asymptotically free gauge theory where the coupling is large. Models such as AdS/QCD have some remnant of the N=4 super Yang-Mills theory left in the description that enforces conformality in the UV (large $r$). That matches the weak coupling physics of these theories - so people bluff and let the dual extend to the far UV. In reality one should cut off the description, eg around 3 GeV or so in QCD, where the coupling becomes weak \cite{Evans:2005ip, Evans:2006ea}. One can simply impose a cut off in the bulk. There is though then a matching problem - one should align the dual to meet QCD in the intermediate coupling period which would mean including some higher dimension operators at the UV boundary. We do now know how to include such operators using Witten's multi-trace prescription \cite{Witten:2001ua}.

\begin{table}
\centering
\caption{The predictions for masses and decay constants (in MeV) for $N_f=2$  SU(3) gauge theory. The columns show the QCD value, the holographic prediction for the theory with massless quarks and the perfected holographic QCD theory with HDOs.  The $\rho$-meson mass has been used to set the scale and other quantities that are used to fix the HDO couplings are indicated by the *s.}
\label{tab-1}       
\begin{tabular}{|c|c|cc|cc|}
\hline
   & QCD & AdS/SU(3) &  & AdS/SU(3) & with\\
       			& 	& 2 F  2 $\bar{F}$ & & 2 F  2 $\bar{F}$ & HDO\\  \hline 
    $M_{\rho}$ 			& 775 			& 775$^*$    &    & 775$^*$  &    \\
    $M_{A}$ 			& 1230 		         & 1183	 & - 4\%& 1230$^*$ &   $g^2_A=5.76149$ \\
    $M_{S}$ 			& 500/990 		& 973  	& +64\%/-2\% & 597  &   $+ 20\%/-40\%$ \\
    $M_{B}$ 			& 938 													& 1451				& +43\% & 938$^*$ &  $g^2_B=25.1558$ \\
    $f_{\pi}$  			& 93 														& 55.6						& -50\%  & 93$^*$ &   $g^2_S= 4.58981$\\
    $f_{\rho}$ 				& 345 													& 321 					& - 7\% & 345$^*$  &    $g^2_V=4.64807$\\
    $f_{A}$ 				& 433 													& 368						& -16\% & 444 &   $+2.5\%$\\   &&&&&\\
    $M_{\rho,n=1}$ 		& 1465 													& 1678					&  +14\% & 1532  &    $+4.5\%$\\
    $M_{A,n=1}$ 		& 1655 													& 1922					& +19\%  & 1789 &   $+8\%$\\
    $M_{S,n=1}$ 		& 990 /1200-1500						& 2009 					&+64\%/+35\%   & 1449  &   $+46\%/0\%$\\
    $M_{B,n=1}$ 		& 1440 													& 2406				& +50\% & 1529 &   $+6\%$\\ \hline
\end{tabular}
\end{table}

Consider adding a Nambu Jona Lasinio operator \cite{Nambu:1961tp} to our description of QCD (Figure 1) \cite{Evans:2016yas,Clemens:2017udk,Jarvinen:2015ofa}
\begin{equation} \label{NJL} \Delta {\cal L} = {g^2 \over \Lambda_{UV}^2} \bar{q}_L q_R \bar{q}_R q_L ~~~~ \rightarrow {g^2 \over \Lambda_{UV}^2} \langle \bar{q}_L q_R \rangle \bar{q}_R q_L ~~~~ \rightarrow m = {g^2 \over \Lambda_{UV}^2} \langle \bar{q}_L q_R \rangle \end{equation}
The logic shown is that in the presence of an NJL term, if the quark condensate forms then there is an effective UV mass. Witten's prescription is to reinterpret the vacuum solutions
 of Figure 1. We now consider solutions that asymptote to a non-zero mass to represent the theory with no bare mass but an NJL term. We read off $m$ and $\langle \bar{q}_L q_R \rangle$ from the solution at the UV cut off and then compute $g^2$ from the relation in (\ref{NJL}).We show the output in the model of two flavour QCD in the lower figure in Figure 1 - there is the usual NJL behaviour - at low couplings only the QCD dynamics generates an IR quark mass but as $g$ goes through a critical value it sharply enhances the dynamical symmetry breaking scale. 

This prescription can be used for all the fields in the Dynamic AdS/QCD action allowing one to include operators $|\bar{q}_L q_R |^2$, $|\bar{q}_L \gamma^\mu q_R |^2$, $|\bar{q}_L \gamma^\mu \gamma_5 q_R |^2$ and $|qqq|^2$. We will not try to match their couplings to perturbative QCD but instead use them as parameters that can be fitted to the low energy meson data - this is like the idea of a perfect action on the lattice \cite{Hasenfratz:1993sp}.

\section{Holographic Description of QCD}

The Dynamic AdS/QCD, holographic model based on the QCD running shown in Figure 1 leads to the predictions shown in Table 1 \cite{Erdmenger:2020flu}. The massless theory compares well to QCD data in that it gets the broad pattern of the masses and couplings correct. However, the pion decay constant is typically low, the baryon mass high and the excited state masses high. The quark and glue content of scalar states is very hard to identify and mixing with glueball states is neglected here, so these identifications are difficult. Including HDOs clearly reduces the predictivity of the theory but also greatly improves the fit including to the radially excited state masses. Including HDOs allows the description to be systematically improved in the spirit of low energy effective field theory.

\section{Beyond QCD - Multi-Flavour Theories}

Recently my collaborators and I have moved to modelling theories with quarks in two different representations of the gauge symmetry  \cite{Erdmenger:2020lvq,Erdmenger:2020flu} . These have been of interest for the construction of composite higgs models with baryons that play the role of top partners \cite{Barnard:2013zea,Ferretti:2014qta}, but here I will just consider them as field theory problems. There has also been some lattice work in this area \cite{Bennett:2019cxd,Ayyar:2018zuk} and I will concentrate on those theories where we have something to compare to.

To model two different representations we simply repeat the Dynamic AdS/QCD model twice with the appropriate runnings of $\gamma/\Delta m^2$ for each of the two condensates we wish to study (and fluctuations of those operator with the various $\gamma$-matrix structures inserted). The only linkage between the sectors we include is that both representations enter the running of the gauge coupling. There is then an interesting question of how the quarks decouple when they go on mass-shell at strong coupling which future lattice work could investigate. We simply remove quarks from the running when they hit their IR mass-shell condition as one would at weak coupling.

\begin{table}
 \caption{  Ground state masses and couplings for the vector, axial-vector, and scalar mesons for the two representations, $F$ and $A_2$, for an Sp(4) and an SU(4) gauge theory. Results are shown from the lattice (Sp(4) from \cite{Bennett:2019cxd}  and SU(4) from \cite{Ayyar:2018zuk} ) and 
 from holography \cite{Erdmenger:2020lvq,Erdmenger:2020flu}.  
  }
 \begin{tabular}{|c||ccc||cc||c|}  \hline
                                                                       & AdS	& AdS 	& Lattice 	& 	Lattice  &  AdS & AdS \\ 
                                                                        &Sp(4)  	& Sp(4) 	& Sp(4) 	& 	 SU(4) &  SU(4) & SU(4) \\ 
     								& $4F,6A_2$ 	 	& quench  		& quench  									& 	$ 4A_2, 2 F, 2 \bar{F} $  & $ 4A_2, 2 F, 2 \bar{F} $ & $ 5A_2, 3 F, 3 \bar{F} $  \\    \hline
    $f_{\pi A_2}$  				  			& 0.120 			& 0.103			& 0.1453(12) 		&  		0.15(4) 				& 0.0997		& 0.111 									\\
    $f_{\pi F}$  							&0.0701 			&  0.0756 		&0.1079(52) 				& 	0.11(2) 					& 0.0953					& 0.109						\\
    $M_{V A_2}$ 		& 1* 					& 1*				& 1.000(32) &  1.00(4) 					&	1*								& 1* 								\\
    $f_{V A_2}$ 						& 0.517 			& 0.518 		& 0.508(18)  		& 	0.68(5)					&	0.489						& 0.516					\\						
    $M_{V F}$ 					   			& 0.814 			& 0.962 		& 0.83(19)			& 	0.93(7) 				&	0.939					& 0.904							\\
   $f_{V F}$ 					   			& 0.364 			& 0.428 		&  0.411(58)	& 		0.49(7) 		 			&	0.461						& 0.491					\\
    $M_{A A_2}$ 				 				& 1.35 				& 1.28 			& 1.75 (13)	& 				 			&	1.37					& 1.32 					\\
    $f_{A A_2}$ 							& 0.520 			& 0.524			& 0.794(70) 			& 						&	0.505				& 0.521											 \\
    $M_{A F}$ 					 			&1.19 				& 1.36 			&  1.32(18) 		& 							&	1.37					& 1.23						\\
    $f_{A F}$ 					 			&0.399 				& 0.462 		& 0.54(11) 			& 					&	0.504			& 0.509 									\\
    $M_{S A_2}$ 				 			& 0.375 			& 1.14			&  			& 			&	0.873			& 0.684 					\\
    $M_{S F}$ 					 			& 0.902 			& 1.25 			&  		& 				&	1.02					& 0.798								\\ \hline
  
 \end{tabular}

\end{table}

\subsection{Sp(4) 4F 6A$_2$}

A first example model is Sp(4) gauge theory with 4 Weyl fermions in the fundamental representation and 6 Weyl fermions in the two-index anti-symmetric representation (of dimension 6). Here there is a lattice study with quenched gauge configurations. We compare to this by using the pure glue running of the Sp(4) theory. The results for the spectrum are shown in Table 2 and require a little digestion.

The first broad point to make about the results is that the predictions for the spectrum of the quenched theory on the lattice and using the Dynamic AdS/QCD model agree well on the pattern of masses and couplings (which are similar to that of QCD). In fact the numerical values here are better than our first attempt at QCD above. Note that the fundamental fermion sector is lighter than the $A_2$ sector, which in the case of the holographic model is because $\gamma$ reaches one first for the $A_2$ representation so the condensation is at a higher scale. 

Given that the holographic model is not a first principles computation, we do not claim to be directly competing with the lattice results - as the lattice groups improve their errors one should look there for the true answer. However, the holographic approach allows us to compute the $\sigma$ meson masses in this model that was beyond the first lattice computation (states with the quantum numbers of the vacuum are hard). Further we can rather simply unquench the holographic model by just including the fermions in the running of $\Delta m^2$ whereas on the lattice this is computationally very expensive - the holographic results are again in Table 2. The holographic model suggest the changes in the spectrum will mostly in this case be at the 10\% level or so except in the $\sigma$ case where as the running slows the $\sigma$ masses fall sharply. The holographic model has therefore contributed additional information that complements the lattice results.

\subsection{SU(4) 3F 3$\bar{ \rm F}$ 5A$_2$}

SU(4) gauge theory with three Dirac fermions in the fundamental representation (or equivalently 3 Weyl fermions in the fundamental and 3 Weyls in the anti-fundamental) and five Weyl fermions in the two-index antisymmetric representation (again a colour sextet) is also a theory of interest for composite higgs models with top partners. Here though for technical reasons the lattice community has had to instead work on the theory with two Dirac fundamentals and four $A_2$s. This lattice simulation is unquenched though. In Table 2 we show both the lattice and Dynamic AdS/QCD results for the spectrum - there's a basic agreement that the $A_2$ sector is heavier and on the spectrum in each of the $A_2$ and fundamental sectors. The holographic model can easily add new information for the axial vector and scalar sectors.

Again though the holographic approach allows us to go further - here we can move to the truly desired fermion content (3 Dirac fermions and 5 $A_2$). The results are shown in Table 2. For the majority of the spectrum the shifts are at below the 10\% level. This is good news because it means the simulations on the lattice are useful, as they would have hoped,  for telling us about the true theory even though they don't quite match its fermion content. Again the holographic model shows a larger shift in the $\sigma$ masses which fall as the running slows. 

\section{Discussion}

We've looked at holographically modelling QCD and two theories with two different representations of fermions. I've tried to argue that the holographic models provide additional insight into the dynamics and spectrum beyond currently achieved  lattice simulations. The main benefit is that one is simply solving some classical differential equations so it is very swift to compute in any given model once the code is set up (you just input the appropriate running for $\gamma$ of the operator you are interested in).  As an example of this, in \cite{Erdmenger:2020flu} we in addition provided spectrum computations for twenty six of the proposed composite higgs models with top partner baryons in \cite{Ferretti:2013kya} (the remaining cases lie in the conformal window at the level of our approximations). A wider survey of theories yet is being constructed on the SCGT Forum web site
which the reader is invited to use and contribute to: \\
\centerline{http://www.southampton.ac.uk/$\sim$evans/SCGT/}

\begin{figure}[h]
    \includegraphics[width=\textwidth]{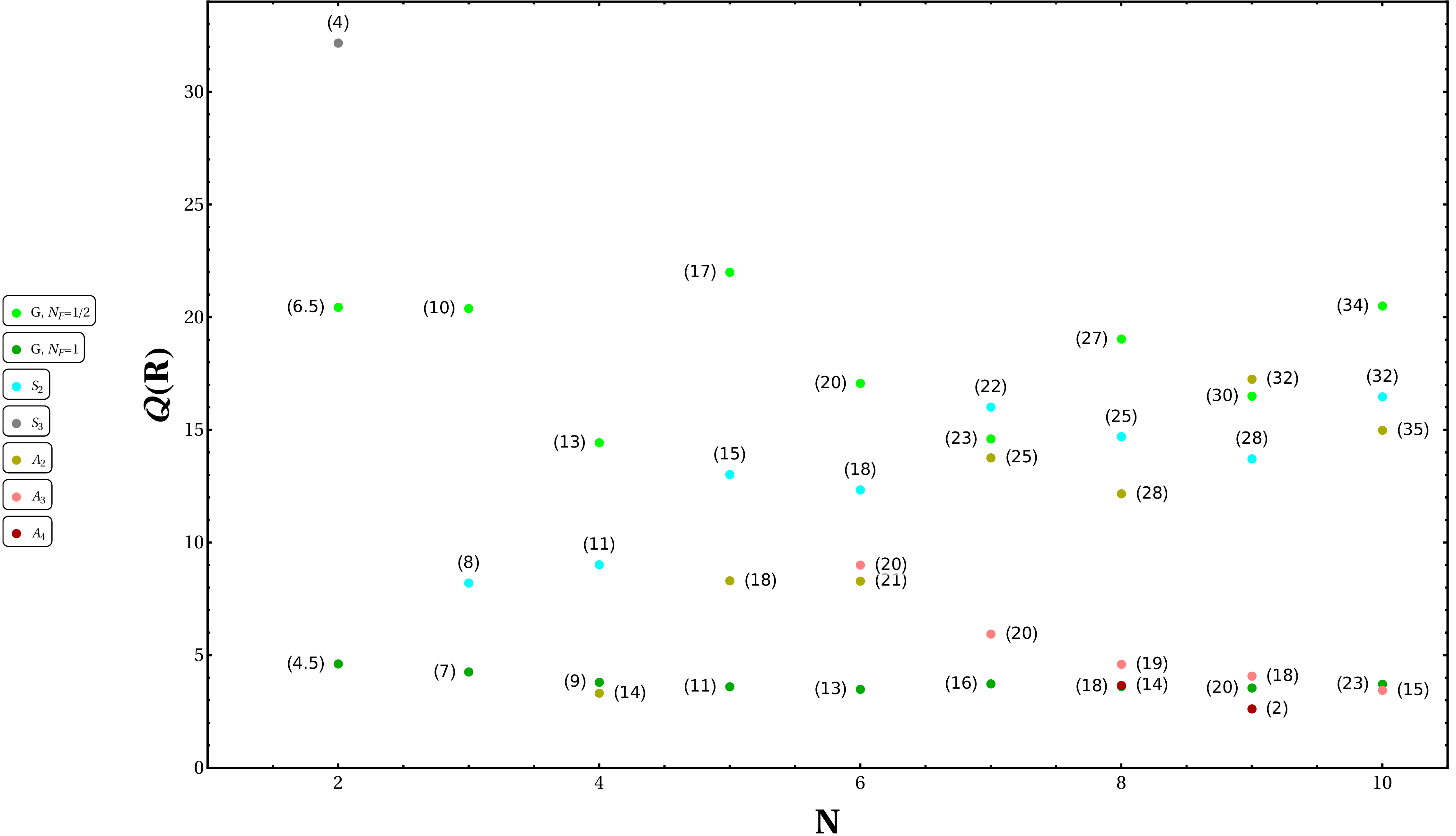}
\caption{
Plot of $ {\mathcal Q}(R) = {\Lambda_{\chi SB ~ R} / \Lambda_{\chi SB ~F} }$ against the number of colours in theories with the minimal number of fermions in the higher dimensional representation (either 1 or 1/2 for real representations) and $N_f^F$ in the fundamental representation. $N_f^F$ has been tuned to maximize $\mathcal{Q}(R)$ and its value is next to the point. We have used red for the fundamental, green for the adjoint, cyan for the rank-2 symmetric, gray for three-rank symmetric, gold for the rank-2 antisymmetric, pink for the $A_3$, maroon for the $A_4$, and blue, orange, black for the R$_{1,2,3}$ respectively. }
\label{fig: plots_sun_2} 
\end{figure}

A further product of modelling these gauge theories is that it throws up interesting dynamical questions that could be answered with future lattice work. One crucial assumption we had to make was that quarks decouple from the dynamics when they go on mass shell at strong coupling in a similar fashion to at weak coupling - if this isn't true the running of these theories will be weaker to lower scales, changing the spectrum.

Another interesting aspect of multi-representation theories is that there is a gap between chiral symmetry breaking scales. If one assumes that the confinement scale lies below the lowest chiral symmetry breaking scale then one could imagine a gap between the highest chiral symmetry breaking scale and confinement. That separation in a full theory would be very interesting. This led Rigatos and I to perform a simple analysis \cite{Evans:2020ztq}, based on perturbative two loop running results,  across all asymptotically free gauge theories to try to maximise the gap. In particular we included the minimum number of Weyl fermions allowed by anomaly constraints of some higher dimension representation $R$ and then added $N_f$ fundamental quarks. We estimated the chiral symmetry breaking scale for $R$ as when $\gamma_R=1$, then integrated out the $R$ representation and ran the coupling with just fundamentals to where $\gamma_F=1$. We then varied $N_f$ to maximize the ratio 
\begin{equation} {\mathcal Q}(R) = {\Lambda_{\chi SB ~ R} \over \Lambda_{\chi SB ~F} } \end{equation} 
ie the ratio of the scales at which the two fermion species condense. We show the results in Figure 2 for a range of SU(N) gauge theories. We can immediately see that there are many theories where $5< Q <20$ which are theories that would have very different phase structure from QCD (for example potentially breaking the chiral symmetry of the R fermions whilst not confining at some finite temperatures). It's worth noting though that these theories are necessarily quite walking since we have dialled the number of fermions to widen the running time gaps between phenomena. They could be quite hard to study on the lattice therefore. There are less walking cases though where one might hope to see different phase structure than QCD although with not such wide gaps. 

In conclusion then, I hope I have successfully argued that holography is a useful tool as we explore the periodic table of gauge theories and their application to beyond the Standard Model physics.

%

\begin{thebibliography}{}
\bibitem{Hill:2002ap}
C.~T.~Hill and E.~H.~Simmons,
Phys. Rept. \textbf{381} (2003), 235-402.

\bibitem{Cacciapaglia:2020kgq}
G.~Cacciapaglia, C.~Pica and F.~Sannino,
Phys. Rept. \textbf{877} (2020), 1-70.

\bibitem{Kribs:2016cew}
G.~D.~Kribs and E.~T.~Neil,
Int. J. Mod. Phys. A \textbf{31} (2016) no.22, 1643004.

\bibitem{Affleck:1983mk}
I.~Affleck, M.~Dine and N.~Seiberg,
Nucl. Phys. B \textbf{241} (1984), 493-534.

\bibitem{Hasenfratz:2019dpr}
A.~Hasenfratz, C.~Rebbi and O.~Witzel,
Phys. Rev. D \textbf{100} (2019) no.11, 114508

\bibitem{Fodor:2016zil}
Z.~Fodor, K.~Holland, J.~Kuti, S.~Mondal, D.~Nogradi and C.~H.~Wong,
Phys. Rev. D \textbf{94} (2016) no.9, 091501.


\bibitem{Erdmenger:2020lvq}
J.~Erdmenger, N.~Evans, W.~Porod and K.~S.~Rigatos,
Phys. Rev. Lett. \textbf{126} (2021) no.7, 071602.

\bibitem{Erdmenger:2020flu}
J.~Erdmenger, N.~Evans, W.~Porod and K.~S.~Rigatos,
JHEP \textbf{02} (2021), 058.

\bibitem{Maldacena:1997re}
J.~M.~Maldacena,
Adv. Theor. Math. Phys. \textbf{2} (1998), 231-252.

\bibitem{Witten:1998qj}
E.~Witten,
Adv. Theor. Math. Phys. \textbf{2} (1998), 253-291.

\bibitem{Erlich:2005qh}
J.~Erlich, E.~Katz, D.~T.~Son and M.~A.~Stephanov,
Phys. Rev. Lett. \textbf{95} (2005), 261602.

\bibitem{DaRold:2005mxj}
L.~Da Rold and A.~Pomarol,
Nucl. Phys. B \textbf{721} (2005), 79-97.

\bibitem{Erdmenger:2007cm}
J.~Erdmenger, N.~Evans, I.~Kirsch and Threlfall,
Eur. Phys. J. A \textbf{35} (2008), 81-133.

\bibitem{Alvares:2012kr}
R.~Alvares, N.~Evans and K.~Y.~Kim,
Phys. Rev. D \textbf{86} (2012), 026008.

\bibitem{Breitenlohner:1982jf}
P.~Breitenlohner and D.~Z.~Freedman,
Annals Phys. \textbf{144} (1982), 249.

\bibitem{Cohen:1988sq}
A.~G.~Cohen and H.~Georgi,
Nucl. Phys. B \textbf{314} (1989), 7-24.

\bibitem{Appelquist:1996dq}
T.~Appelquist, J.~Terning and L.~C.~R.~Wijewardhana,
Phys. Rev. Lett. \textbf{77} (1996), 1214.

\bibitem{Alho:2013dka}
T.~Alho, N.~Evans and K.~Tuominen,
Phys. Rev. D \textbf{88} (2013), 105016.

\bibitem{Jarvinen:2011qe}
M.~Jarvinen and E.~Kiritsis,
JHEP \textbf{03} (2012), 002.


\bibitem{Evans:2005ip}
N.~Evans, J.~P.~Shock and T.~Waterson,
Phys. Lett. B \textbf{622} (2005), 165-171.

\bibitem{Evans:2006ea}
N.~Evans and A.~Tedder,
Phys. Lett. B \textbf{642} (2006), 546-550.

\bibitem{Witten:2001ua}
E.~Witten,
[arXiv:hep-th/0112258 [hep-th]].

\bibitem{Nambu:1961tp}
Y.~Nambu and G.~Jona-Lasinio,
Phys. Rev. \textbf{122} (1961), 345-358.


\bibitem{Jarvinen:2015ofa}
M.~Jarvinen,
JHEP \textbf{07} (2015), 033

\bibitem{Evans:2016yas}
N.~Evans and K.~Y.~Kim,
Phys. Rev. D \textbf{93} (2016) no.6, 066002.

\bibitem{Clemens:2017udk}
W.~Clemens and N.~Evans,
Phys. Lett. B \textbf{771} (2017), 1-4.



\bibitem{Hasenfratz:1993sp}
P.~Hasenfratz and F.~Niedermayer,
Nucl. Phys. B \textbf{414} (1994), 785-814.

\bibitem{Barnard:2013zea}
J.~Barnard, T.~Gherghetta and T.~S.~Ray,
JHEP \textbf{02} (2014), 002.

\bibitem{Ferretti:2014qta}
G.~Ferretti,
JHEP \textbf{06} (2014), 142.

\bibitem{Bennett:2019cxd}
Bennett, D.~K.~Hong, J.~W.~Lee, C.~J.~D.~Lin, B.~Lucini, M.~Mesiti, M.~Piai, J.~Rantaharju and D.~Vadacchino,
Phys. Rev. D \textbf{101} (2020) no.7, 074516.

\bibitem{Ayyar:2018zuk}
V.~Ayyar, T.~Degrand, D.~C.~Hackett, W.~I.~Jay, E.~T.~Neil, Y.~Shamir and B.~Svetitsky,
Phys. Rev. D \textbf{97} (2018) no.11, 114505.

\bibitem{Ferretti:2013kya}
G.~Ferretti and D.~Karateev,
JHEP \textbf{03} (2014), 077.

\bibitem{Evans:2020ztq}
N.~Evans and K.~S.~Rigatos,
Phys. Rev. D \textbf{103} (2021), 094022.



\end{thebibliography}
%
%

\end{document}